\documentclass{iaus}
\usepackage{graphicx}

\title[Structure of dense cores in HMSF regions] 
{Physical and chemical structure of dense cores in regions of high
mass star formation}

\author[Zinchenko et al.]   
{Igor Zinchenko$^1$, Lev Pirogov$^1$, Paola Caselli$^2$, \break
Lars E.B. Johansson$^3$, Sergey Malafeev$^1$ \break \and Barry Turner$^4$}

\affiliation{$^1$Institute of Applied Physics, Russian Academy of Sciences,
\break Ulyanov str. 46, Nizhny Novgorod 603950, Russia
\break email: zin@appl.sci-nnov.ru, pirogov@appl.sci-nnov.ru,
 malafeev\_s@mail.ru \\[\affilskip]
$^2$Arcetri Astrophysical Observatory, Largo E. Fermi, 5, 50125 Firenze, Italy
\break email: caselli@arcetri.astro.it \\[\affilskip]
$^3$Onsala Space Observatory, S-43992, Onsala, Sweden
\break email: leb@oso.chalmers.se \\[\affilskip]
$^4$National Radio Astronomy Observatory,  520 Edgemont Road, \break
 Charlottesville, VA 22903-2475, USA
\break email: bturner@nrao.edu
}

\pubyear{2005}
\volume{227}  
\pagerange{xxx--yyy}
\date{?? and in revised form ??}
 \jname{Massive Star Birth: A Crossroads of
Astrophysics} \editors{R. Cesaroni, E. Churchwell, M. Felli \&
C.M. Walmsley, eds.}
\begin{document}

\maketitle

\begin{abstract}
We found that in regions of high mass star formation the CS
emission correlates well with the dust continuum emission and is
therefore a good tracer of the total mass while the N$_2$H$^+$
distribution is frequently very different. This is opposite to
their typical behavior in low-mass cores. The behavior of other high density
tracers varies from source to source but most of them are closer
to CS. Radial density profiles in massive cores are fitted by
power laws with indices about $-1.6$, as derived from the dust
continuum emission. The radial temperature dependence on
intermediate scales is close to the theoretically expected one for
a centrally heated optically thin cloud. The velocity dispersion
either remains constant or decreases from the core center to the
edge. Several cores including those without known embedded IR
sources show signs of infall motions. They can represent the
earliest phases of massive protostars. There are implicit
arguments in favor of small-scale clumpiness in the cores.

\keywords{astrochemistry, stars: formation, ISM: abundances, ISM:
molecules, radio lines: ISM}
\end{abstract}

\firstsection 
\section{Introduction}
Many important aspects of high mass star formation (HMSF) remain
unknown or poorly known. This is caused by the much more
complicated appearance of HMSF regions in comparison with their
low-mass counterparts and by the relative difficulties of their
studies due to larger distances, etc. Therefore, detailed
investigations of the structure and kinematics of HMSF cores are
of great importance. For example, different models of star
formation predict different density profiles in the cores and
different frequency of outflow occurrence.

An important question in this respect is a selection of reliable
tracers of physical parameters in HMSF regions. 
Chemical effects can seriously
distort their appearance. It is now well established that the
central parts of dense low mass cloud cores suffer strong
depletion of many molecules (in particular, CO, HCO$^+$ and CS)
onto dust grains (\cite[Caselli, Walmsley, 
Tafalla, \etal\ 1999]{Caselli99}; \cite[Caselli, Walmsley, Zucconi,
\etal\ 2002]{Caselli02};
\cite[Kramer, Alves, Lada, \etal\ 1999]{Kramer99}; 
\cite[Willacy, Langer, \& Velusamy 1998]{Willacy98};
\cite[Jessop \& Ward-Thompson 2001]{Jessop01}; 
\cite[Tafalla, Myers, Caselli, \etal\ (2002)]{Tafalla02}; 
\cite[Bergin, Ciardi, Lada, \etal\ 2001]{Bergin01}), suggesting that
CS (so far considered a high density tracer) does not actually
probe the central core regions. On the other hand, N$_2$H$^+$ is
usually an excellent tracer of dust continuum emission 
(\cite[Caselli \etal\ 2002]{Caselli02}), implying that this species depletes at
higher densities compared to CO, despite of the same binding
energies (\cite[\"Oberg, van Broekhuizen, Fraser, \etal\ 2005]{Oberg05}).

The situation is less clear in HMSF cores where the mechanisms of
chemical differentiation are probably different (e.g. not linked
to molecular freeze-out). There were several studies which show
chemical variations in a few similar regions 
(\cite[Ungerechts, Bergin, Goldsmith, \etal\ (1997)]{Ungerechts97}; 
\cite[Bergin, Goldsmith, Snell, \etal\ 1997]{Bergin97}). Here,
we present observations of a statistically significant sample of
HMSF cores.

\section{Chemical differentiation in HMSF regions}\label{sec:chemdiff}
A comparison of our CS 
(\cite[Zinchenko, Mattila \& Toriseva 1995]{Zin95};
\cite[Zinchenko, Pirogov \& Toriseva 1998]{Zin98};
Pirogov \etal, in preparation) and N$_2$H$^+$ 
(\cite[Pirogov, Zinchenko, Caselli, \etal\ 2003]{Pirogov03}) 
maps in many cases shows striking differences
between them. Our further studies of dust continuum emission and
other high density tracers (HCN, HNC, HCO$^+$, their isotopes,
etc.) reveal certain relationships between them. The details of
these observations are presented elsewhere (Pirogov \etal, in
preparation, Zinchenko \etal, in preparation). Here we summarize
the main results.

\begin{figure*}[htb]
\begin{minipage}{0.48\textwidth}
\centering
\resizebox{\hsize}{!}{\rotatebox{-90}{\includegraphics{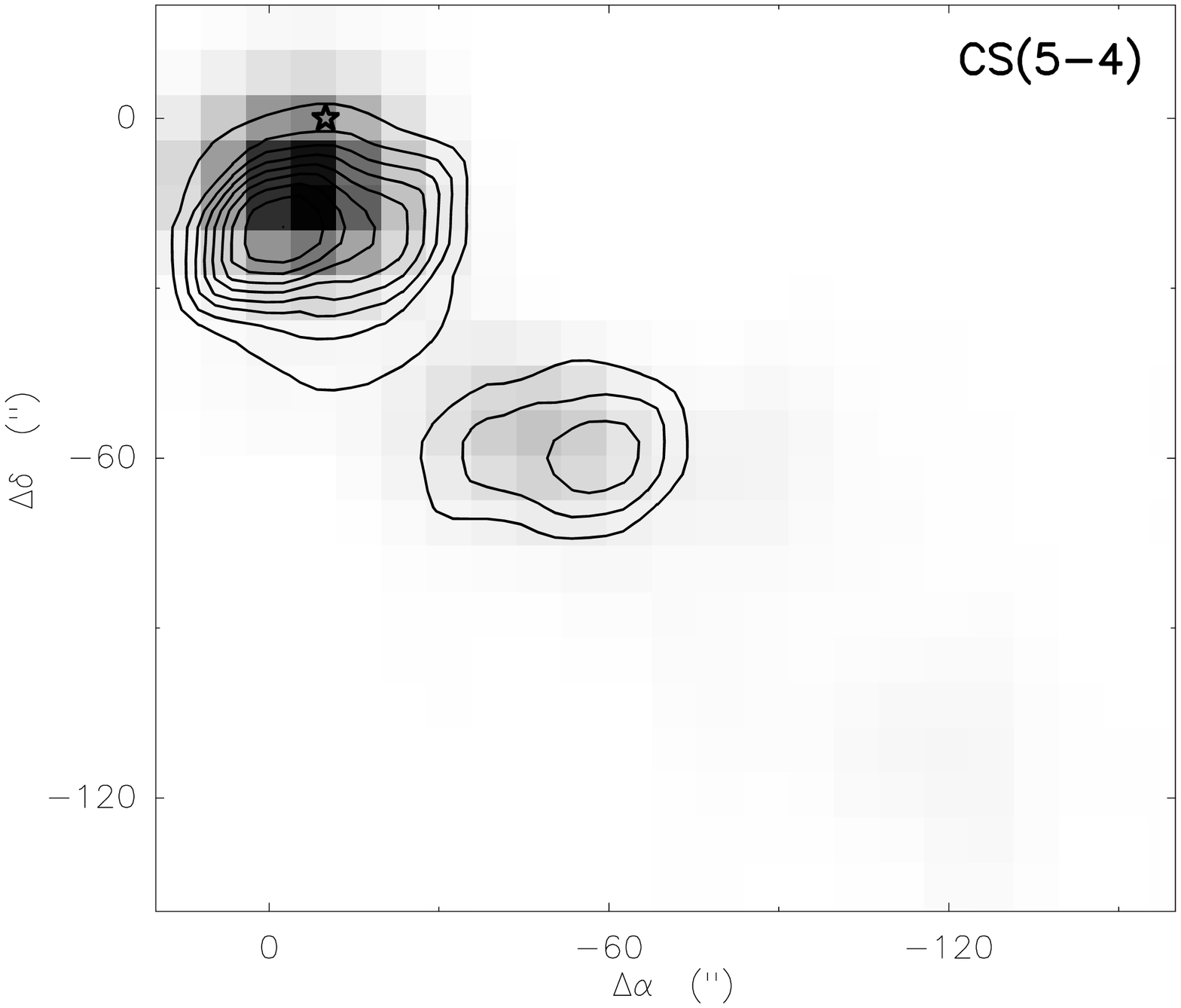}}}
\end{minipage}
\hfill
\begin{minipage}{0.48\textwidth}
\centering
\resizebox{\hsize}{!}{\rotatebox{-90}{\includegraphics{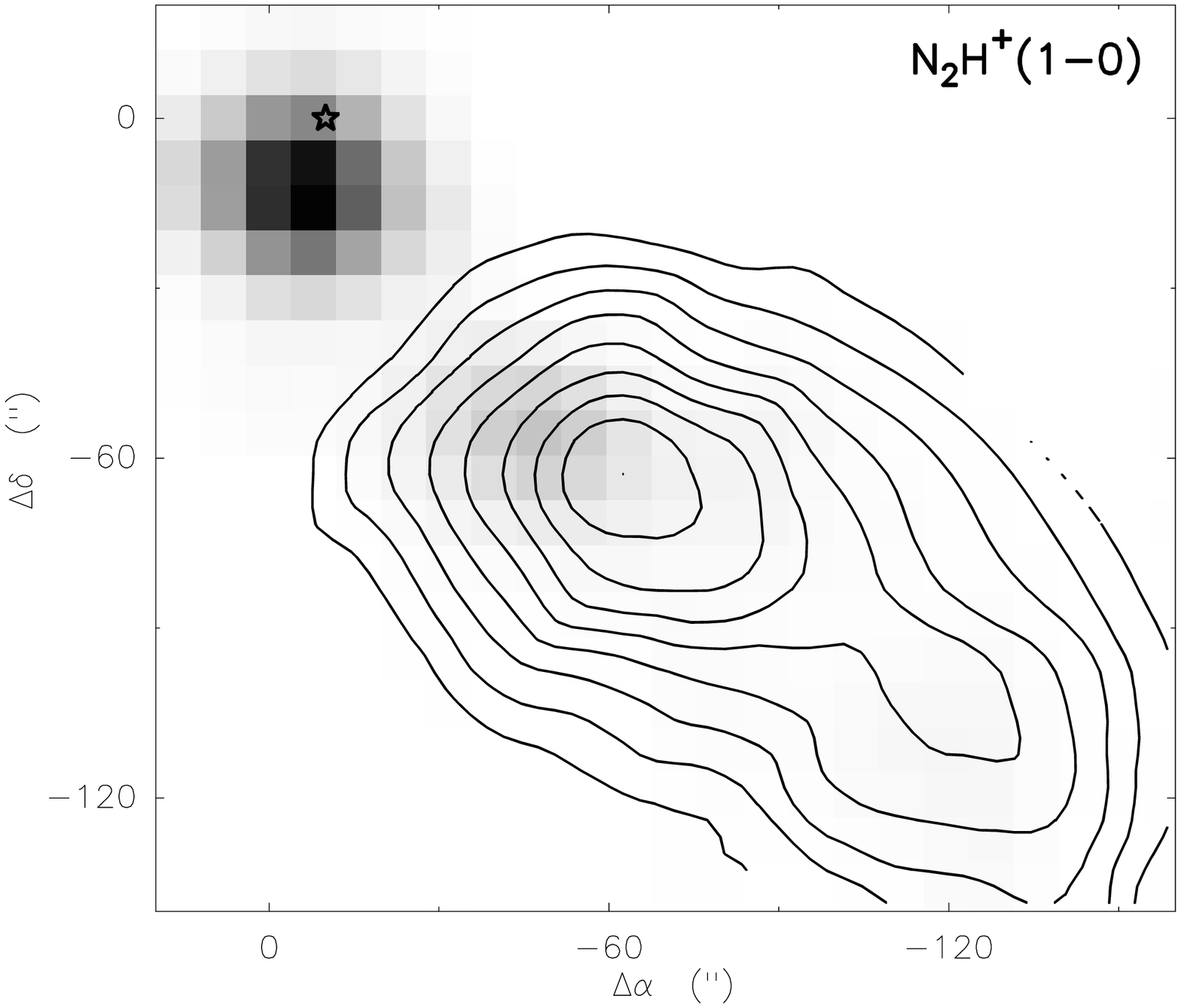}}}
\end{minipage}
\caption{Maps of CS(5--4) and N$_2$H$^+$(1--0) emission (contours)
overlaid on the 1.2 mm dust continuum map (grey scale) in
G285.26-0.05. Star marks the position of the IRAS point source.
All maps are obtained at SEST.} \label{fig:g285}
\end{figure*}

First, we found a good correlation between CS and dust emission in
most cases. At the same time N$_2$H$^+$ distributions are
frequently very different. An example of such comparison is shown
in Fig.~\ref{fig:g285}. The millimeter wave dust (and CS) emission
peaks are also usually associated with strong FIR sources (in
particular, from the IRAS catalogue). On the other hand the
N$_2$H$^+$/CS ratio usually reaches the highest values towards
clumps without such association.

This behavior is opposite to that typical for cold low-mass cores.
The variations of the CS/N$_2$H$^+$ intensity ratio reach more than
an order of magnitude.
It is worth noting that the N$_2$H$^+$ optical depth is usually small
(\cite[Pirogov \etal\ 2003]{Pirogov03}) 
while the CS optical depth can be rather
high. This means that the saturation effects can only decrease the
apparent variations of the CS/N$_2$H$^+$ ratio.

\begin{figure*}[htb]
\begin{minipage}[b]{0.19\textwidth}
\centering
\resizebox{\hsize}{!}{\rotatebox{-90}{\includegraphics{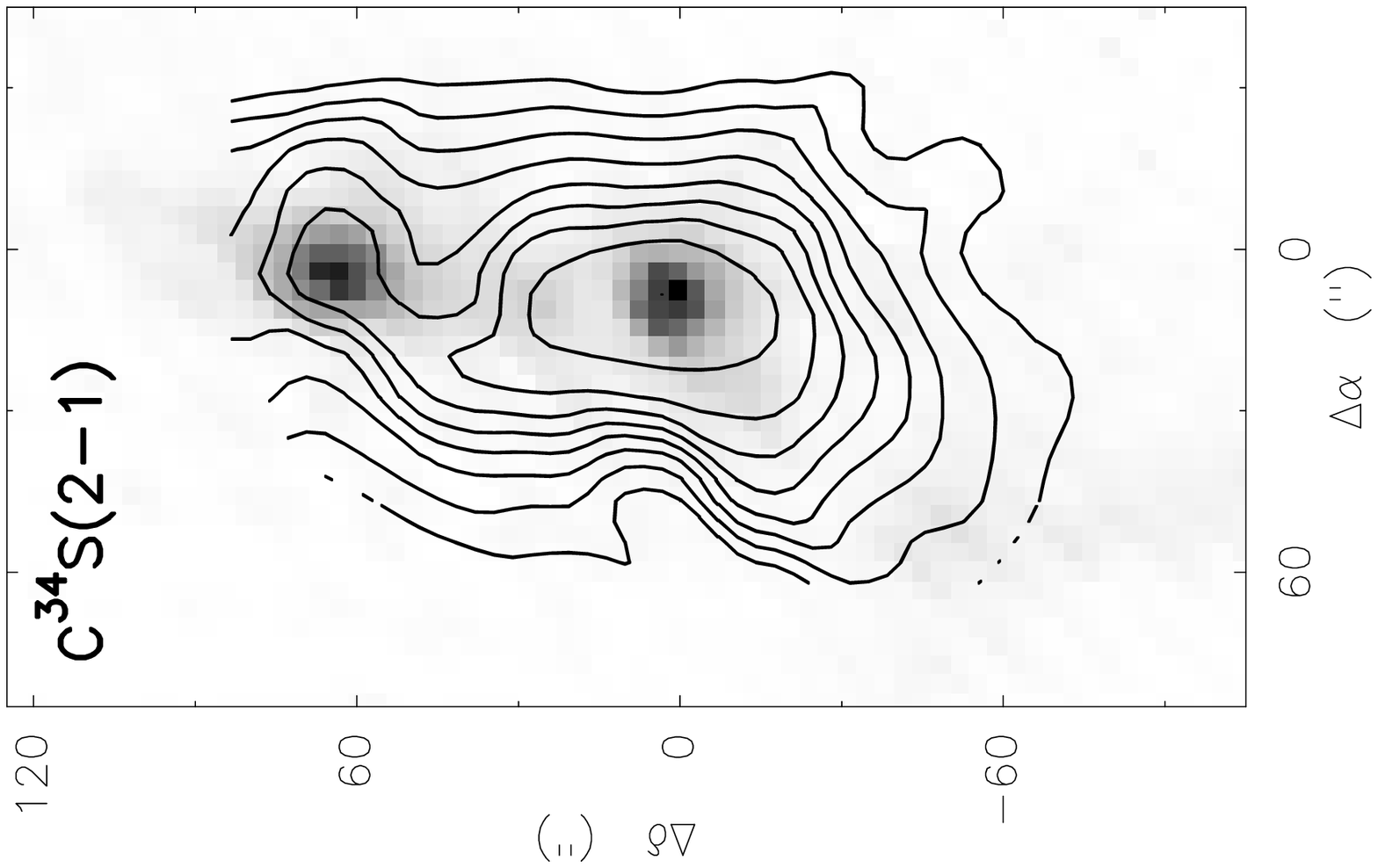}}}
\end{minipage}
\hfill
\begin{minipage}[b]{0.19\textwidth}
\centering
\resizebox{\hsize}{!}{\rotatebox{-90}{\includegraphics{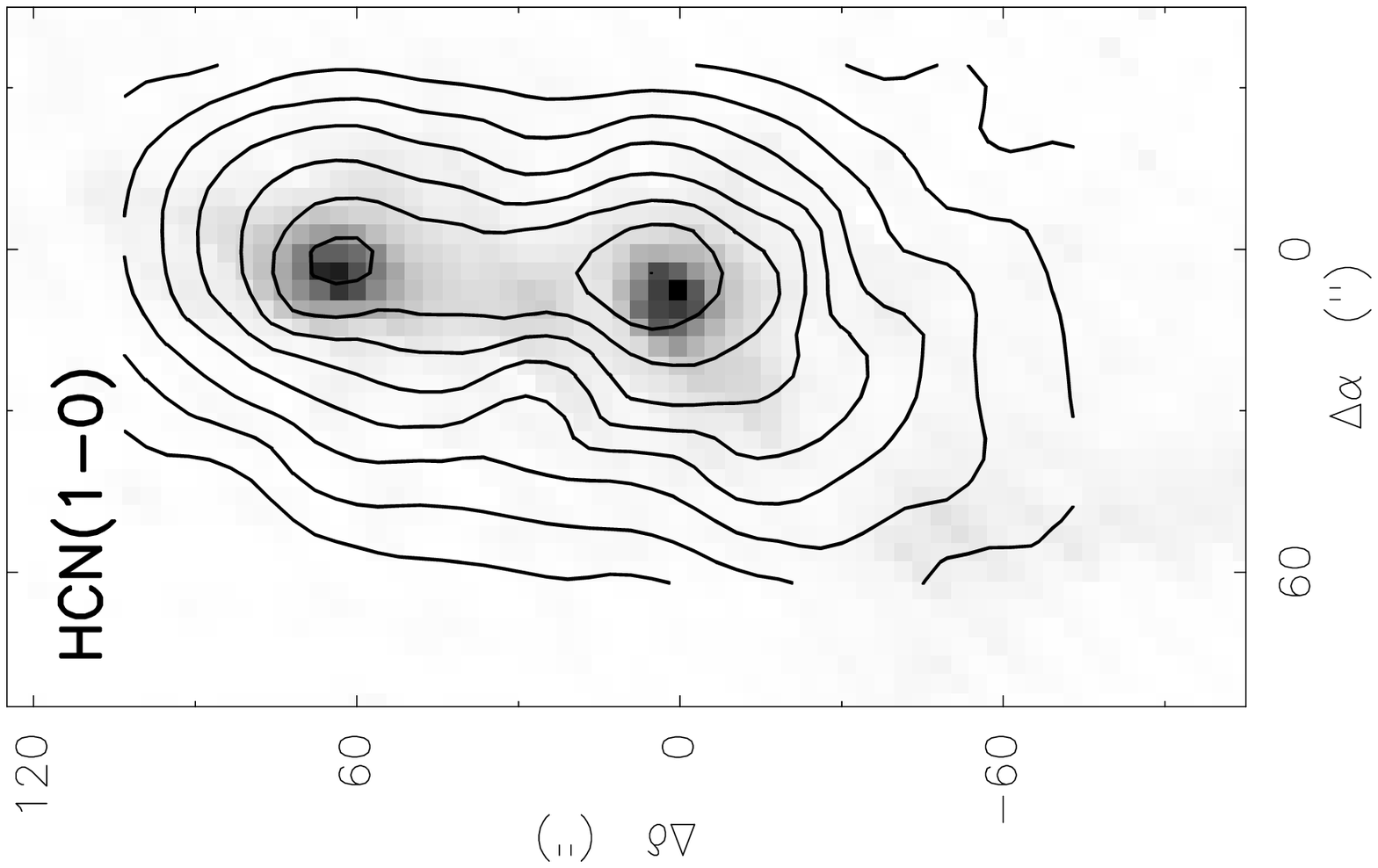}}}
\end{minipage}
\hfill
\begin{minipage}[b]{0.19\textwidth}
\centering
\resizebox{\hsize}{!}{\rotatebox{-90}{\includegraphics{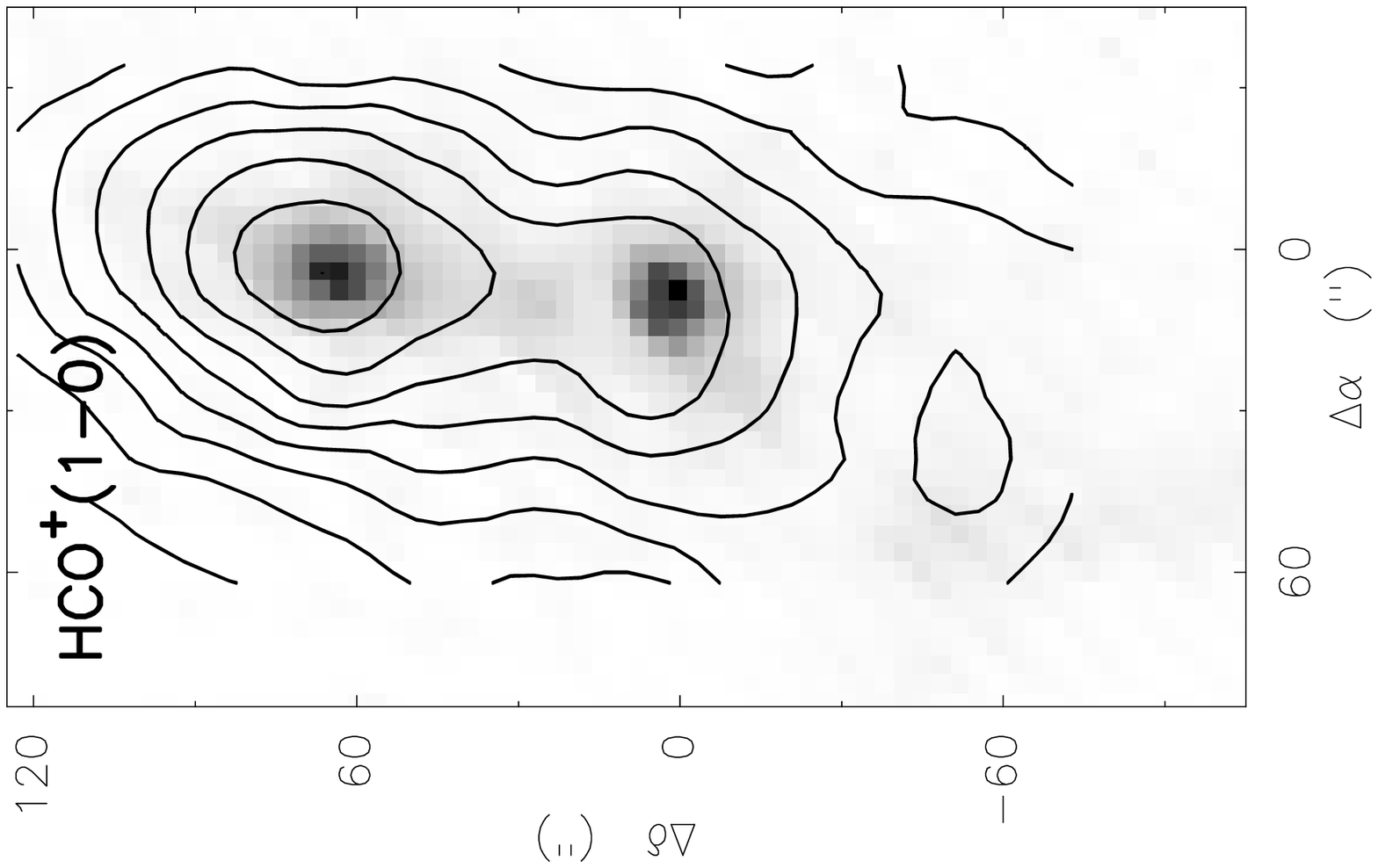}}}
\end{minipage}
\hfill
\begin{minipage}[b]{0.19\textwidth}
\centering
\resizebox{\hsize}{!}{\rotatebox{-90}{\includegraphics{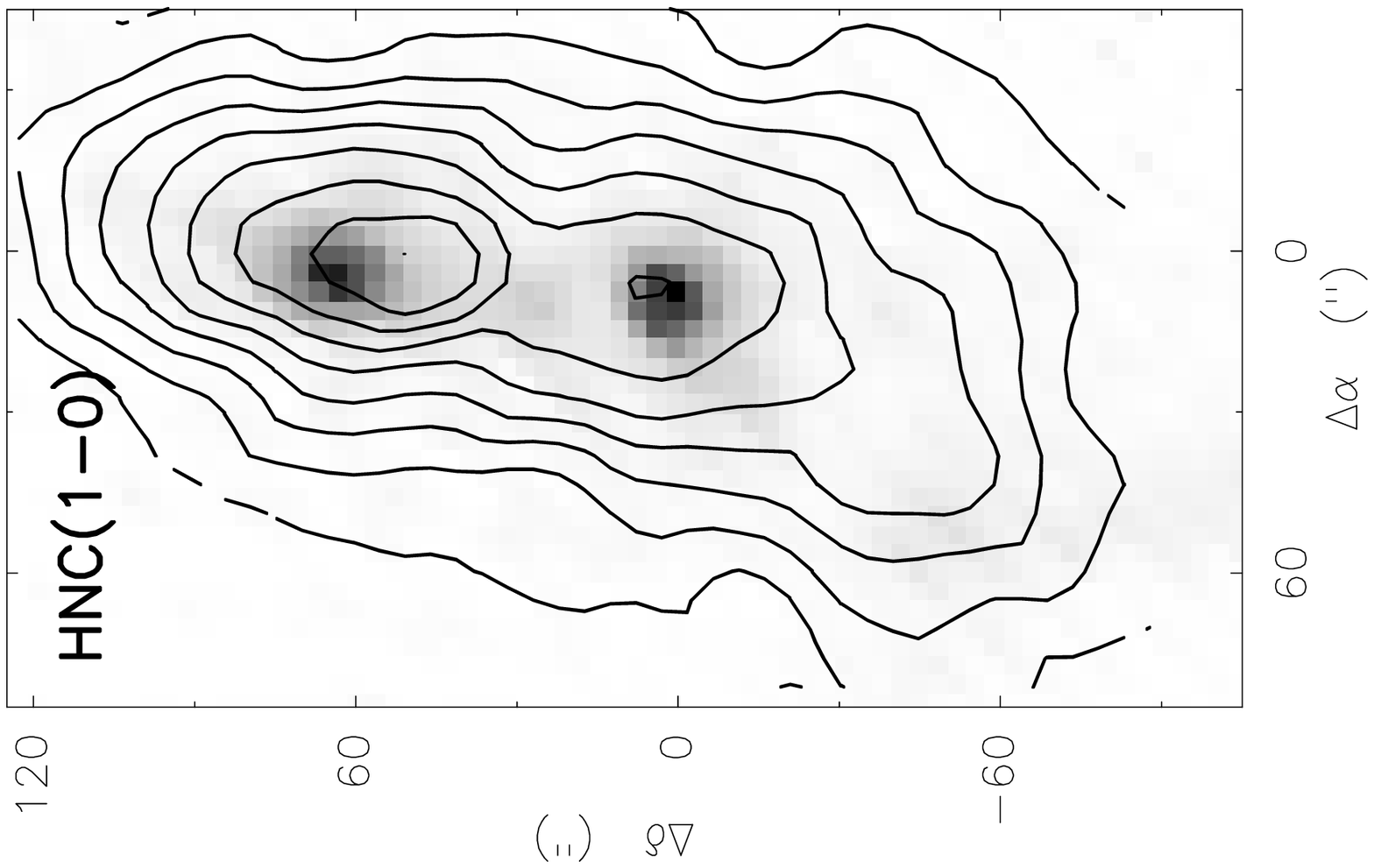}}}
\end{minipage}
\hfill
\begin{minipage}[b]{0.19\textwidth}
\centering
\resizebox{\hsize}{!}{\rotatebox{-90}{\includegraphics{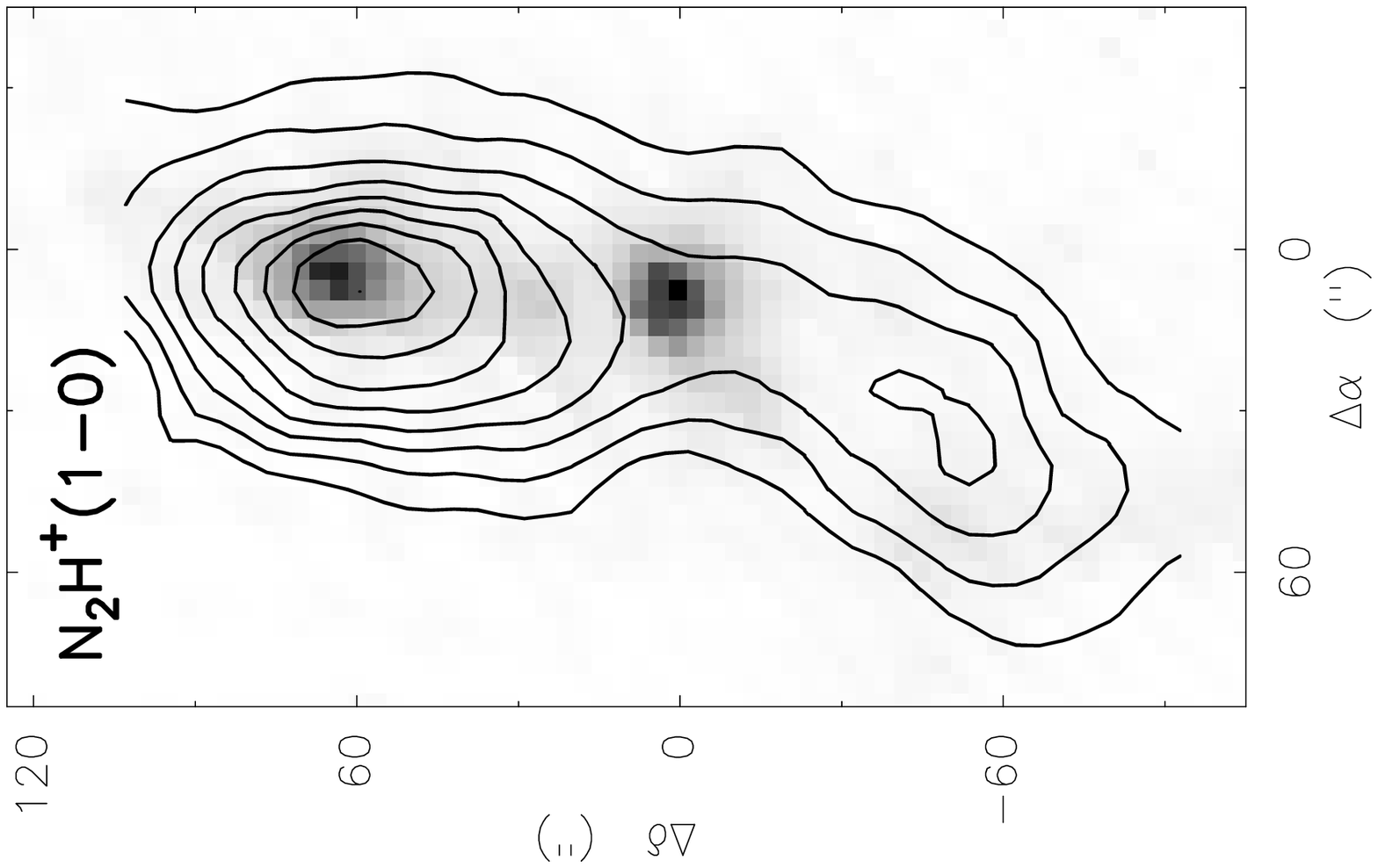}}}
\end{minipage}
\caption{Maps of S255 in various molecular lines (contours) obtained
with the OSO-20m radio telescope overlaid on
the map of 1.2 mm dust continuum emission (grayscale) obtained
by us with the IRAM-30m radio telescope.
}
\label{fig:s255}
\end{figure*}

The behavior of other high density tracers varies from source to
source but most of them are closer to CS. An exception is HNC
which is an intermediate case in this sense between CS and
N$_2$H$^+$. In Fig.~\ref{fig:s255} we present as an example maps
of several high density tracers in S255 overlaid on the dust
continuum map. It shows two continuum peaks almost equal in
intensity. The nature of these two components is different. The
central one is associated with a luminous cluster of IR sources,
whereas toward the northern one an ultracompact H~{\sc ii} region
(G192.58-0.04) was detected. CS and HCN follow the dust
distribution quite well, while N$_2$H$^+$ is very different.
HCO$^+$ map in S255 looks similar to N$_2$H$^+$ but in some other
sources it is closer to CS. The same can be said about ammonia.
A comparison of the C$^{18}$O and dust continuum data for this source
shows no sign of CO freeze out which suggests that the dust temperature
is above 20~K, the sublimation temperature of CO.

We studied variations of kinetic temperature in these sources by
CH$_3$C$_2$H observations.
In most cases no significant temperature difference between the CS
and N$_2$H$^+$ peaks was found 
(\cite[Malafeev, Zinchenko, Pirogov, \etal\ 2005]{Malafeev05}). 
Typical temperatures which we derived for these peaks are $\sim 30-40$~K. 
E.g. in S255
we found kinetic temperatures of $\sim 40$~K for both peaks
mentioned above. Therefore, the observed chemical differentiations
cannot be explained by molecular freeze-out as in low mass cores.

Density differences between the CS and N$_2$H$^+$ peaks are also
not very significant as follows from our density estimates based
on CS and CH$_3$OH LVG modelling. At the same time the gas
densities at CS peaks are somewhat higher than at the N$_2$H$^+$
ones and in general there is a correlation between CS emission
intensity and gas density.

One possible explanation for the observed chemical
differentiations was proposed by 
\cite[Lintott, Viti, Rawlings, \etal\ (2005)]{Lintott05}. 
They suggested that the enhancement of
CS/$\mathrm{N_2H^+}$ abundance ratio may be
related to the high dynamical activity in these regions which
could enhance the rate of collapse of cores above the free-fall
rate. Consequently, high gas densities would be achieved before
freeze-out had removed the molecules responsible for the
$\mathrm{N_2H^+}$ loss, while the high densities promote CS
formation. Our data show that in most cases the line widths at the
CS peaks are somewhat larger than at the N$_2$H$^+$ peaks which is
consistent with this model although in S255 the reverse is true.

\section{Internal structure and kinematics of HMSF cores}
\label{sec:radial}

\subsection{Radial density profiles}
\begin{figure*}[htb]
\begin{minipage}{0.48\textwidth}
\centering
\resizebox{\hsize}{!}{\includegraphics{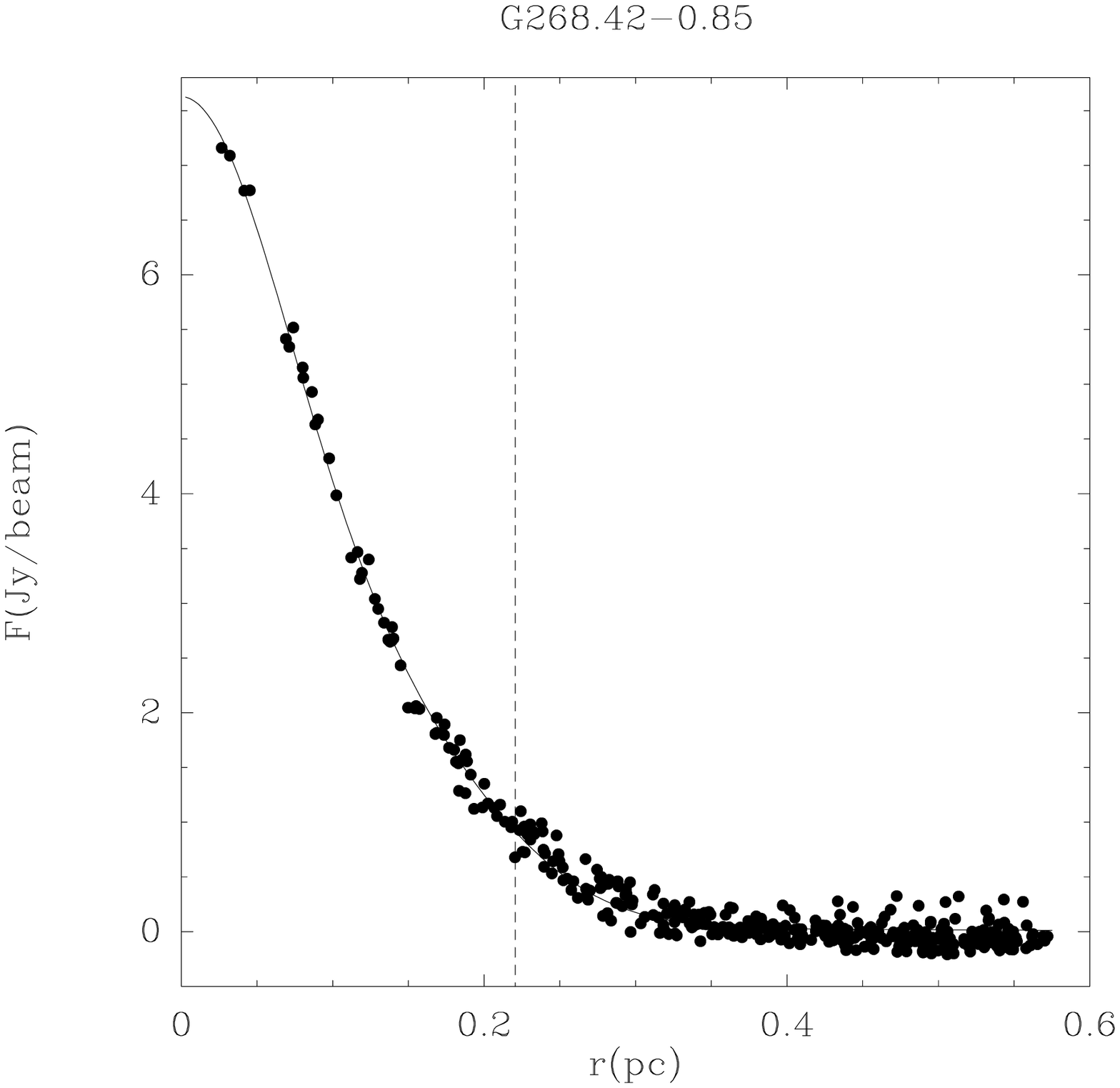}}
\end{minipage}
\hfill
\begin{minipage}{0.48\textwidth}
\centering
\resizebox{\hsize}{!}{\includegraphics{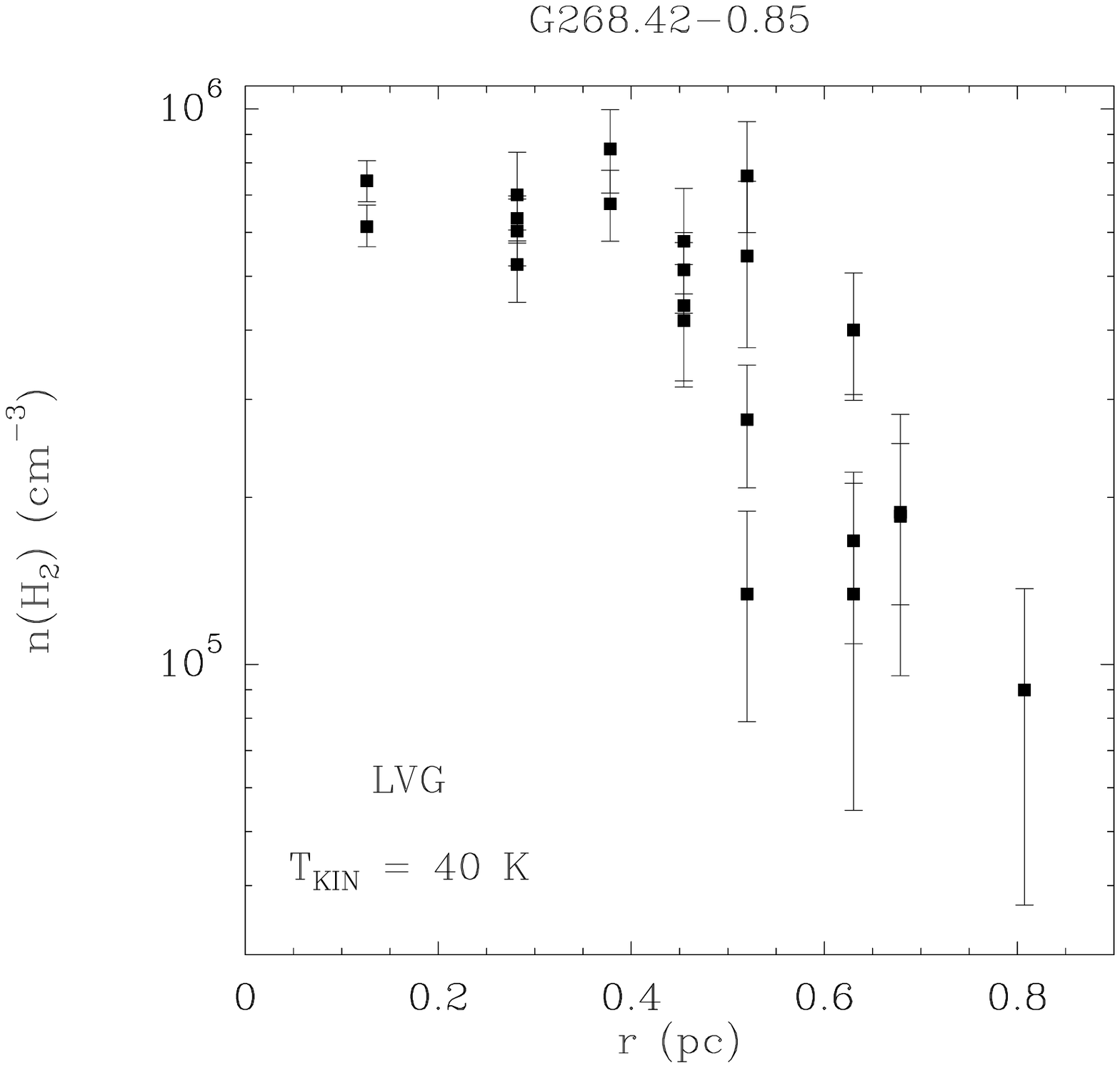}}
\end{minipage}
\caption{An example of the radial dependence of dust emission intensity
fitted by a power law function convolved with a gaussian beam (left panel)
and the radial dependence of density derived from LVG analysis
of the CS $J=2-1$ and $J=5-4$ lines (right
panel).} \label{fig:rdens}
\end{figure*}

We studied the density profiles in the cores using the molecular
line and dust continuum maps. From the N$_2$H$^+$ data we derived
the $\sim r^{-2}$ density profile assuming constant N$_2$H$^+$
excitation and abundance along the radius 
(\cite[Pirogov \etal\ 2003]{Pirogov03}). 
However, this assumption is not justified and
probably more reliable conclusions can be drawn from dust
observations. The analysis of the dust continuum maps was similar.
We fitted the 2D intensity distribution for nearly circular clumps
with a convolution of a power law function with a gaussian beam.
The details of this analysis are presented elsewhere (Pirogov
\etal, in preparation). An example of this fitting is shown in
Fig.~\ref{fig:rdens}.

In total, the data for 17 clumps obtained at SEST were analyzed.
The power law index is $p=-1.0\pm 0.1$ for clumps with embedded
IRAS point sources and $p=-0.6\pm 0.1$ for clumps without such
sources. Assuming temperature gradient in the former sources
(described by a power law index of $-0.4$, see below) and
isothermal latter sources, this corresponds to $\sim r^{-1.6}$
density profiles for both types of clumps.

Such density profiles are in agreement with several other studies
(e.g. \cite[Beuther, Schilke, Menten, \etal\ 2002]{Beuther02}; 
\cite[Mueller, Shirley, Evans II, \etal\ 2002]{Mueller02}). 
They rule out some theoretical models (e.g.
so-called ``logatropic'' one) but are consistent with the
``standard'' theory of star formation (\cite[Shu 1977]{Shu77}).

It is worth noting that these dependencies are obtained for
\emph{mean} volume densities which are derived from column
densities. Densities found for regions of molecular emission from
excitation analysis are usually significantly higher and have
different dependence on radius (as shown in
Fig.~\ref{fig:rdens}). This can be a consequence of small scale
clumpiness in these regions. The volume filling factor should be
$\leq 0.4$.

\subsection{Radial temperature profiles}
\begin{figure*}[htb]
\begin{minipage}{0.45\textwidth}
\centering
\resizebox{\hsize}{!}{\rotatebox{-90}{\includegraphics{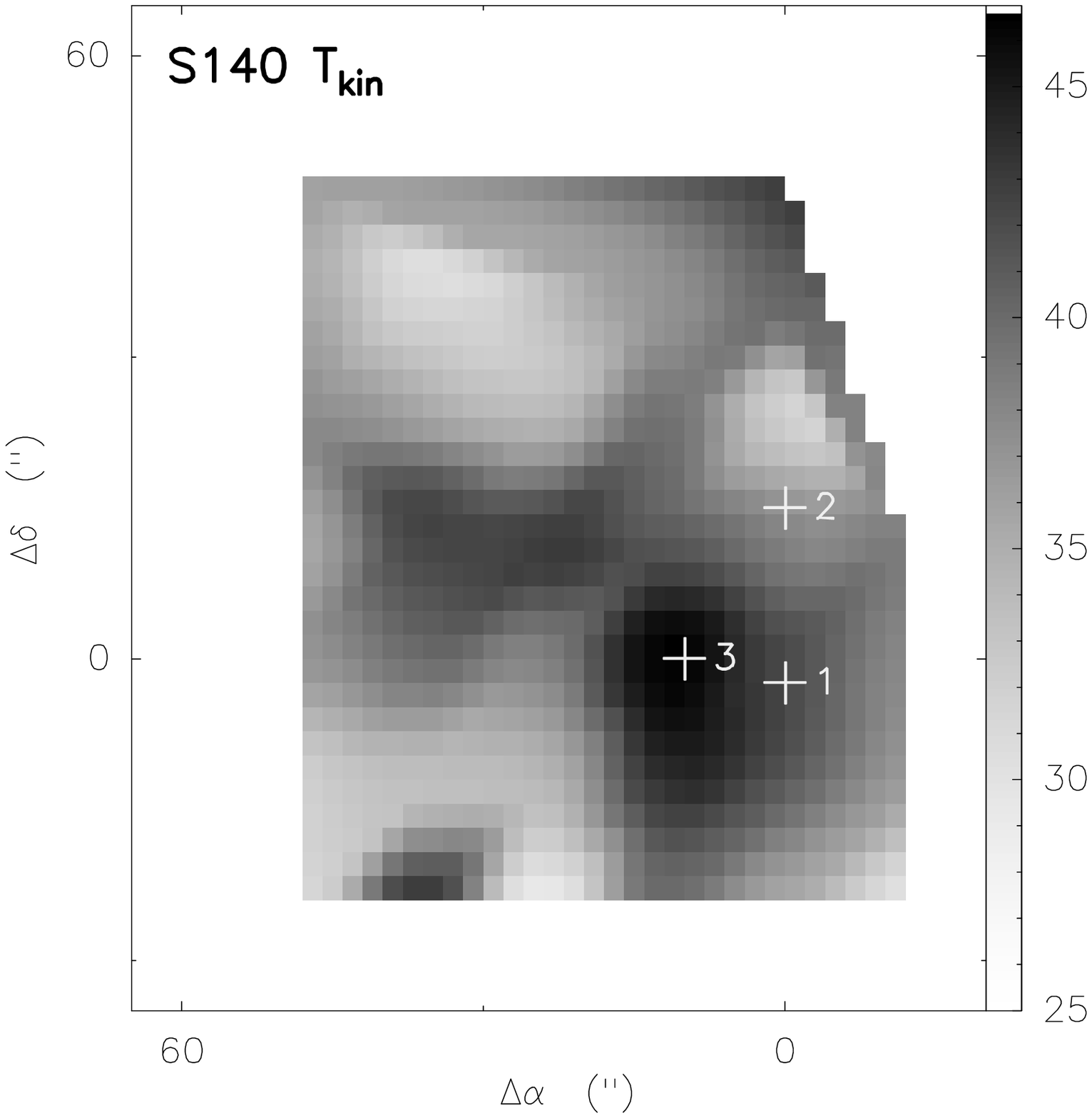}}}
\end{minipage}
\hfill
\begin{minipage}{0.52\textwidth}
\centering
\resizebox{\hsize}{!}{\includegraphics{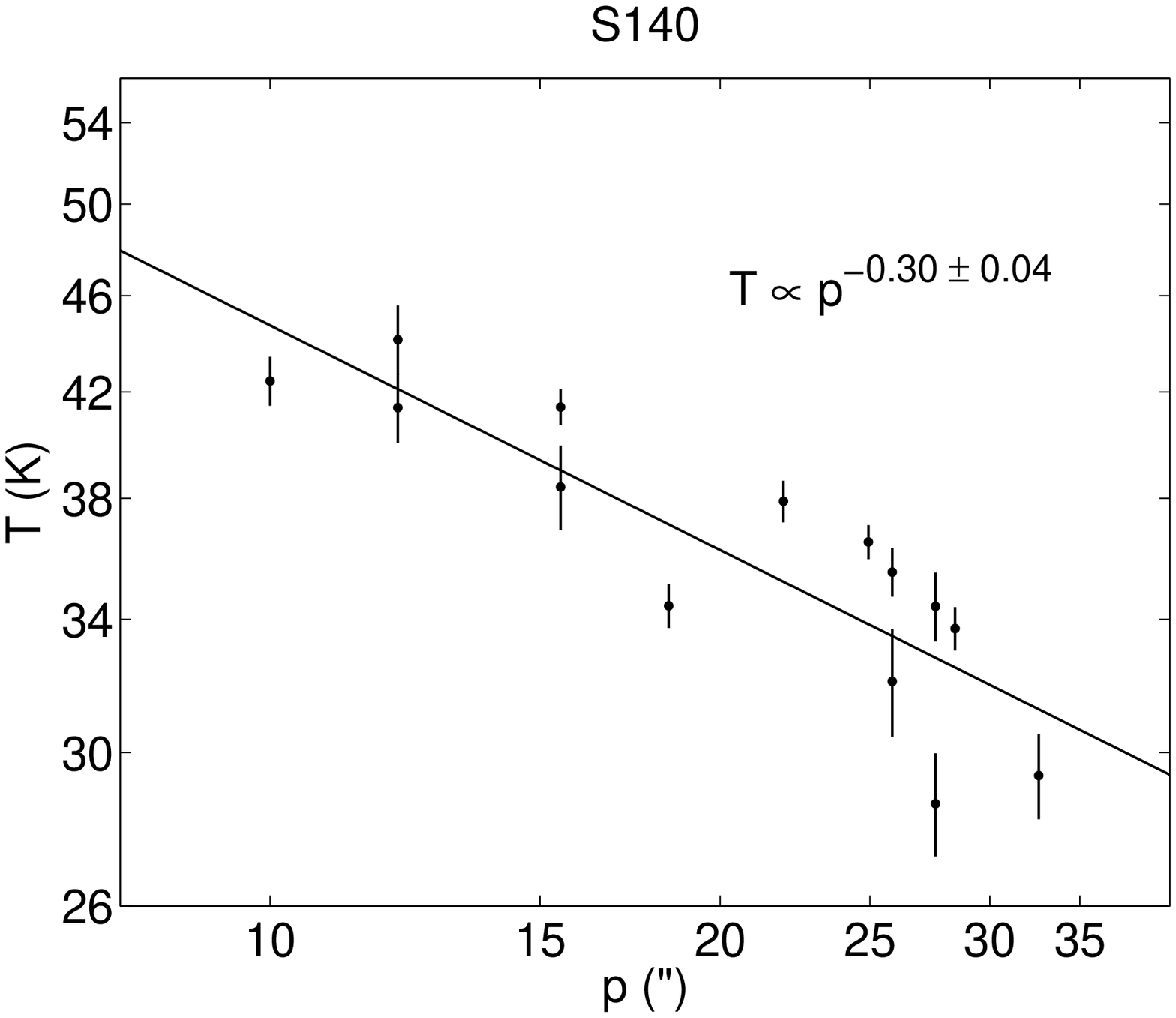}}
\end{minipage}
\caption{Distribution of kinetic temperature in S140 as derived
from the CH$_3$C$_2$H $J=13-12$ data obtained in 2004 with the
IRAM-30m radio telescope (left panel) and the dependence of this
temperature on the projected distance from the clump center for
the strongest clump (right panel). Positions of IRS 1, 2 and 3 
(\cite[Evans II, Mundy, Kutner, \etal\ 1989]{Evans89}) are marked. 
Apparent variations of kinetic
temperature at the periphery of the map are unreliable and are
probably caused by low signal to noise ratio.}
\label{fig:s140-tkmap}
\end{figure*}

We studied the temperature distribution in the cores on the basis
of high quality CH$_3$C$_2$H $J=13-12$ maps obtained at IRAM 30-m
telescope (the beam width at this frequency is $\sim
12^{\prime\prime}$). An example of temperature map (for S140) is
shown in Fig.~\ref{fig:s140-tkmap}. These maps could be
deconvolved into several distinct clumps and the dependence of
temperature on the projected distance from the clump center was
analyzed. An example of this dependence for the main clump in S140
is also shown in Fig.~\ref{fig:s140-tkmap}.

These dependencies were compared with model calculations of
temperatures which would be derived for optically thin
CH$_3$C$_2$H emission from a spherically symmetric cloud with a
power law temperature and density gradients (CH$_3$C$_2$H was
assumed to be thermalized at densities $n>10^4$~cm$^{-3}$). We
found that the radial temperature dependence in these clumps can
be fitted by a power law with indices between --0.3 and --0.4.

This dependence is in agreement with the theoretically expected
one for a centrally heated optically thin cloud. Some other recent
studies found somewhat different dependencies. E.g. 
\cite{Fontani02} derived steeper temperature gradients from
comparison of temperatures found from different species and at the
same time argued for rather constant temperature in regions of
CH$_3$C$_2$H emission. The results presented here do not support
these conclusions.

\subsection{Radial velocity dispersion profiles}
Our CS and N$_2$H$^+$ data show that the velocity dispersion in
HMSF cores either remains constant or decreases outwards
(\cite[Zinchenko 1995]{Zin95a}; 
\cite[Lapinov, Schilke, Juvela, \etal\ 1998]{Lapinov98}; 
\cite[Pirogov \etal\ 2003]{Pirogov03}).
A plausible explanation could be found in a higher degree of
dynamical activity of gas in central regions of HMSF cores,
including differential rotation, infall motions and turbulence due
to winds and outflows from massive stars. It is worth noting that
\cite{Caselli95} derived an opposite trend from
comparison of line widths and sizes of the emission regions for
different species in the same source. The reason for this
discrepancy is not clear yet.

\subsection{Core rotation and associated outflows}
Many HMSF cores mapped in N$_2$H$^+$ show systematic velocity
gradient fields, implying nearly uniform rotation 
(\cite[Pirogov \etal\ 2003]{Pirogov03}). 
There is a correlation ($cc=0.9$) between
direction angle of total velocity gradient and elongation angle.
This fact justifies an assumption that elongation of cores and
clumps could be due to rotation. However, the ratio of rotational
to gravitational energy is only $\sim 0.01$ on the average, so
that rotation should not play a significant role in core dynamics.

The frequency of occurrence of high velocity outflows in the
studied sample was found to exceed $\sim 40$\% using SO line wings
as an indicator (\cite[Zinchenko 2002]{Zin02}) which is in agreement
with other available estimates.

\subsection{Infall motions and massive protostars}
In several cases molecular line profiles (especially HCO$^+$) show
features indicative of infall motions (red-shifted
self-absorption). This is a case both for clumps with and without
strong embedded IR sources. However, the latter ones are probably
at an earlier stage of evolution, and they are characterized by a
relatively high N$_2$H$^+$/CS ratio, as mentioned above. Thus, one
can speculate that this ratio can be one of the indicators of the
earliest phases of massive protostars.

\section{Conclusions}\label{sec:concl}
We have shown that effects of chemical differentiation in warm
massive cores are very different from those in cold low mass
cores. In particular, CS is a good tracer of total mass here,
while N$_2$H$^+$ is not. A relatively high N$_2$H$^+$/CS ratio can indicate
the earliest phases of massive protostars. The internal structure
of HMSF clumps is more or less consistent with the standard theory
of star formation.

\begin{acknowledgments}
The work was supported by
Russian Foundation for Basic Research grant 03-02-16307
 (in part) and INTAS grant 99-1667 (in part).
The research has made use of the SIMBAD database,
operated by CDS, Strasbourg, France.
\end{acknowledgments}





\end{document}